# Insights from Graph Theory on the Morphologies of Actomyosin Networks with Multilinkers

Yossi Eliaz[1,2], Francois Nedelec[3], Greg Morrison[1,2], Herbert Levine[2,4], and Margaret S. Cheung[1,2,5]*

[1]*Department of Physics, University of Houston, Houston, Texas 77204, USA*

[2]*Center for Theoretical Biological Physics, Rice University, Houston, Texas 77005, USA*

[3]*Sainsbury Laboratory, Cambridge University, Bateman Street, CB2 1LR Cambridge, UK*

[4]*Department of Physics, Northeastern University, Boston, Massachusetts 02115, USA*

[5]*Department of Bioengineering, Rice University, Houston, Texas 77030, USA*

*Corresponding author: mscheung@uh.edu

# ABSTRACT

Quantifying the influence of microscopic details on the dynamics of development of the overall structure of a filamentous network is important in a number of biologically relevant contexts, but it is not obvious what order parameters can be used to adequately describe this complex process. In this paper, we investigated the role of multivalent actin-binding proteins (ABPs) in reorganizing actin filaments into higher-order complex networks via a computer model of semiflexible filaments. We characterize the importance of local connectivity among actin filaments as well as the global features of actomyosin networks. We first map the networks into local graph representations and then, using principles from network-theory order parameters, combine properties from these representations to gain insight on the heterogeneous morphologies of actomyosin networks at a global level. We find that ABPs with a valency greater than two promote filament bundles and large filament clusters to a much greater extent than bivalent multilinkers. We also show that active myosin-like motor proteins promote the formation of dendritic branches from a stalk of actin bundles. Our work motivates future studies to embrace network theory as a tool to characterize complex morphologies of actomyosin detected by experiments, leading to a quantitative understanding of the role of ABPs in manipulating the self-assembly of actin filaments into unique architectures that underlie the structural scaffold of a cell relating to its mobility and shape.

# I. INTRODUCTION

Living cells actively regulate the morphologies of actomyosin networks to control their shape and mechanical forces during various cellular processes [1,2]. They do so, in part, by regulating the activity of specific actin-binding proteins (ABPs). Some ABPs, such as Myosin II (a molecular motor) [3] depend on the hydrolysis of ATP to actively produce force within actomyosin networks. Other ABPs, e.g., α-actinin, are passive and coordinate contraction and change the morphology of actomyosin networks by cross-linking two filaments [4]. There also exists another kind of ABPs, such as calcium/calmodulin-dependent kinase II (CaMKII) [5,6], that can link together more than two filaments simultaneously [7] to assemble actin filaments [7,8]. In addition to its multivalent actin-binding nature, CaMKII may translate chemical calcium signals into mechanical responses in actomyosin networks when it binds to a calcium-bound calmodulin (CaM) protein and consequently dissociates from actin filaments [8]. The dynamic connectivity between actin filaments facilitates certain transformations, often linked to the function of these networks, such as expansion cell motility [9,10], or contractility [8,11-13], or morphological plasticity of dendritic spines in neuronal cells [14].

Because ABPs regulate the dynamic connectivity between actin filaments, they reorganize the topology of actomyosin networks; thus, its architecture becomes increasingly complex. CaMKII stands out among other bivalent ABPs due to its ability to bind more than two filaments at once [7]. CaMKII is a large protein complex with 12 identical subunits assembling into a double-decked hexagon [15]. Each subunit has pieces of modules to bind CaM as well as actin filaments. Although CaMKII has 12 possible binding sites with actin filaments, only a fraction of these binding sites are permitted to associate with actin filaments such as junctions and bundles due to volume exclusion [7,16]. We hypothesize that the actin-binding multivalency of a single protein complex, such as CaMKII, facilitates hierarchal mesoscopic actin scaffolds by increasing the possible local linking combinations of actin filaments.

We tested our hypothesis that linkers' multivalency drives the complexity of network architecture using Cytosim [17], a software developed to simulate mesoscopic cytoskeletal biological systems [11,18]. Cytosim represents polymeric actin as filaments with bending elasticity and models linkers as stochastic entities that can bind and connect filaments into a network.

Extending Cytosim's modeling of motors, actin filaments and bivalent ABPs (crosslinkers), we incorporated a model for multivalent ABPs (multilinkers). The simulations of actomyosin networks with multilinkers allowed us to investigate how the valency of multilinkers shapes random actomyosin networks from random mesh grids into a hierarchical scaffold.

To quantitatively describe the emergent complexity in the architecture of this network from the ABP-mediated assembly of actin filaments, we reduced its high dimensionality using a wide range of order parameters drawn from polymer physics theory, gelation theory, and graph/network theory. We organized all of these order parameters based on their correlations, and plotted the time evolution of selected order parameters, to better understand the quantitative features of the topology that can be captured by individual parameters. We find that the order parameters from network theory provide useful qualitative insights for investigating the properties of a filament network connected by ABPs, which are not captured by polymer or gelation order parameters. By treating an actin filament as a node, the actomyosin network forms a graph generated by their spatial distributions, with the formation of cliques, clusters, or communities in varying graininess. A key measure from network theory is to characterize the hierarchy among nodes by assessing the "importance" of each node in the network using a variety of "centrality" measures drawn from the literature [19,20]. The measure of centrality, such as betweenness [21], distinguishes the topographical importance of actin filaments (or nodes) in a global network. Additional order parameters, such assortativity [22], characterize the correlations between centrality within the network, a quantity not easily captured using traditional polymer physics and gelation theory. Such expanded measures signify how topological complexity of an actomyosin network emerges from multivalent linkages of an ABP. Furthermore, we establish testable predictions by generating a state diagram in planes of motor activity and multilinker concentration based on observations of the emergent structural elements of the network. With this expanded toolbox of quantitative descriptors, we gained new insight by observing heterogenous structures beyond a typical two-state narrative (e.g. sol-gel), as gelation theory or polymer physics theory enables.

### III. MODELS AND METHODS

#### A. Coarse-grained model of actomyosin networks

We used a coarse-grained computational model to study the morphology and structure of actomyosin networks. In our 3D model, an actin filament is represented as an incompressible bendable polar fiber with rigidity of 0.075 pNμm² (persistence length of 18 μm) [23]. A myosin motor is a bivalent ABP with two walking heads, each of which operates independently and walks toward the plus end of a different filament. Additionally, we have two types of passive linker species: one that resembles an α-actinin bivalent crosslinker and second, the multivalent linker inspired by CaMKII with a variable valency that is fixed along the simulation. In this article, for brevity, we refer a multivalent linker as a 'multilinker'. As depicted in **Fig. 1**, the simulated systems contain filaments, passive bivalent crosslinkers, passive multivalent linkers (multilinkers), and active bivalent motors.

## B. Collective Constrained Langevin dynamics

The filaments and multilinkers are represented by N three-dimensional vertices, which are combined into a vector $\mathbf{x}(t)$ of dimension 3N, describing the physical objects in the system at time t. The equations of motion are evolved by Langevin dynamics via the vectorial stochastic differential equation [17]: $d\mathbf{x}(t) = \boldsymbol{\mu}\, \mathbf{f_{tot}}(\mathbf{x}, t)dt + d\mathbf{B}(t)$, where $\boldsymbol{\mu}$ is a 3N × 3N diagonal matrix consisting of the mobility coefficients of all objects, $\mathbf{f_{tot}}(\mathbf{x}, t)$ is a vector contains all the forces acting on the points $\mathbf{x}(t)$ at time t. Lastly, $\mathbf{B}(t)$ is a vector that recapitulates the random Brownian noise due to molecular collisions; its $i$th coordinate, $\mathbf{B}_i(t)$ is a temperature-dependent value, drawn from a normal distribution with a zero mean and a standard deviation equals to $\sqrt{2dt\mathbf{D}_i}$, where $dt$ is the integration time. Per Einstein's relation we set the diffusion coefficient of the $i$th random molecular degree of freedom to be $\mathbf{D}_i \equiv \boldsymbol{\mu}_{ii} k_B T$, where $k_B$ is Boltzmann's constant and $T$ is the fixed temperature of the system.

## C. Motor activity and binding and unbinding events

We used the software Cytosim [17] to propagate the equations of motion. We augmented its feature to include multilinkers (https://gitlab.com/f.nedelec/cytosim/-/tags/multilinkers) and an example input file is provided in the Supplementary Information. After the collective Brownian mechanics have been calculated, Cytosim executes two sub-routines to account for chemical processes such as binding, unbinding, and motor walking. The first subroutine simulates the binding and unbinding events of actin-binding proteins according to fixed rates $k_{\mathrm{on}}$ and $k_{\mathrm{off}}$, respectively (please see the Supplementary Information Method Section). The second subroutine emulates motor activity on the filaments, based on the forces experienced by motors (please see the Supplementary Information Method Section), displacing the motors along the filaments [17].

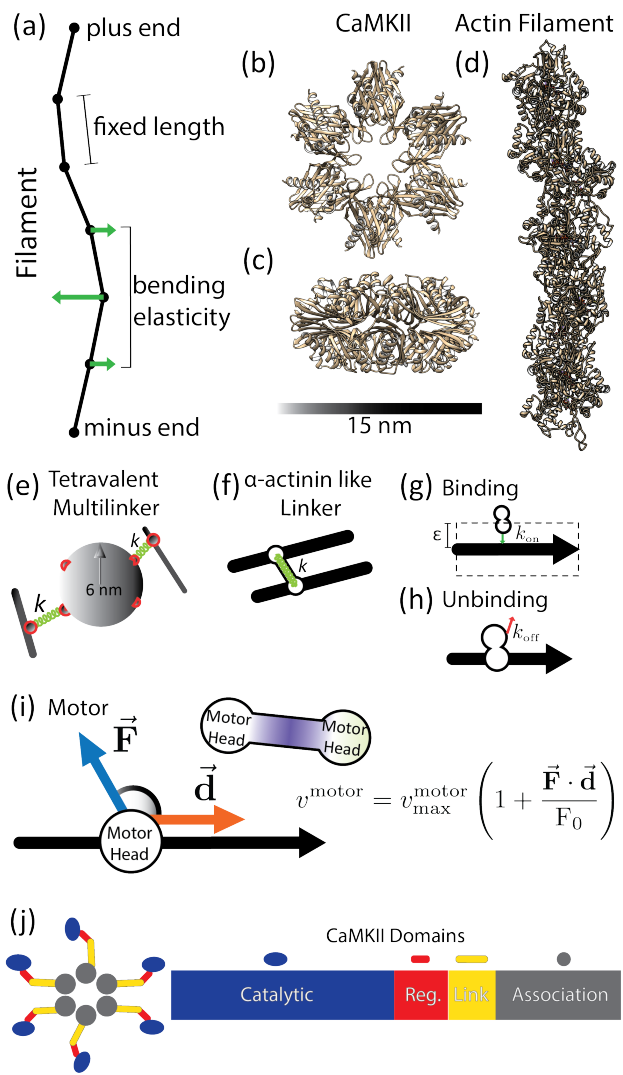

**FIG. 1**. Illustration of Cytosim's model and its components. A vertex-based model of actin filament in (a) with an accompanying crystal structure of 10 globular actin (G-actin) monomers build up a filamentous actin (F-actin) in (d), based on PDB ID: 3J8I. In (b) and (c) are the side and front views of CaMKII crystal structure (PDB ID: 3SOA) visualizing only the association domains of CaMKII. The protein complexes in (b), (c), and (d) were generated using UCSF Chimera [24]. A multilinker of $v$-valency was built using a sphere with $v$ active sites arranged on its surface such as to maximize the minimum distance between any two binding entities. The special case of tetravalent ($v = 4$) linker with a 6 nm radius attached to two filaments is depicted here (e). An α-actinin-like bivalent crosslinker is modeled as a spring between two filaments as shown in (f). In (g) and (h), there is an illustration of binding and unbinding of a diffusive particle such as: a motor,

a crosslinker, or a multilinker from a filament. A myosin-like motor, walking on a filament and exerting a force, is depicted in (i) with the motor's force-velocity relationship. CaMKII's functional domains are shown in (j) including the association domain that binds F-actin [8] and the catalytic domain that binds to ATP and is regulated by the regulatory (Reg.) domain, which becomes active by calcium/calmodulin signaling [25].

### D. Simulation Parameters

Our system is a 1μm$^3$ cube with an internal medium viscosity of 0.5 pNμm$^{-2}$s. Filaments have length of 0.5 μm and are uniformly distributed in the cubic system at $t = 0$. The α-actinin-like crosslinkers are modeled as a Hookean spring between two binding entities where each binding entity has a binding range of 17.5 nm. The myosin-like motors are inextensible objects with two independent motor heads separated by a resting length of 100 nm. Multilinkers are a third type of ABPs to emulate CaMKII's multivalency. A multilinker of valency $v$ has $v$ binding hands residing on the surface of a sphere of radius 6 nm, and all hands are linked to the sphere with $v$ springs.

To study the effect of the multilinkers' valency on the global network we have run simulations at varied conditions. Each of these conditions is defined by five parameters: the number multilinkers in the system $N_{\text{multilinkers}} \in \{250, 500, 1000\}$, their valencies $v \in \{2,3,4,5,6,7\}$, the number of motors $N_{\text{motors}} \in \{250, 500, 1000\}$, the number of filaments $N_{\text{filaments}} \in \{250, 500, 1000\}$, and the number of α-actin-like crosslinkers $N_{\text{crosslinkers}} \in \{10, 250, 500, 1000\}$). For example, one system included 250 pentavalent ($v = 5$) multilinkers, 500 filaments, 1000 motors, and 10 crosslinkers.

In the results section we extensively compare between two system conditions: "high motor content", i.e., the ratio among motors, multilinkers, α-actinin crosslinkers and filaments in the 1 $\mu m^3$ box with 1,000 filaments is $1:1:1:1$, and systems at "low motor content" condition, i.e., the ratio among motors, multilinkers, α-actinin crosslinkers and filaments in the 1 $\mu m^3$ box with 1,000 filaments is $1:100:100:100$. Using these two sets, "low" and "high" motor content, we compare the behavior of different order parameters. We have conducted 600-second-long mesoscopic simulations using Cytosim, and for the "low" and "high" motor content, we collect statistics from 30 simulations with different random initial conditions. Moreover, in a few cases we highlight three specific multivalency conditions $v = 2, 3$, and 6 to study how multilinker's valency tunes the complexity of actomyosin networks at both "low" and "high" motor content.

## E. Order Parameters

Here, we explore the use of various order parameters to analyze actomyosin network simulations. An order parameter should reflect a physically meaningful feature of a system, ideally reducing the high-dimensional configuration space of an actomyosin network into a single quantity. We employed three interrelated families of order parameters: (i) those based on the locations of the coarse-grained geometrical vertices of actin filaments; (ii) those based on the interconnectivity of between filaments by ABPs; and (iii) those based on graph theory and the graph representation of actomyosin network as an undirected graph, which is a mathematical structure to represent nodes and their connections. We keep the details and physical meaning of most measures from (i) and (ii) in the Supplementary Information, and provide the details from (iii) below.

*E.1 Network theory order parameters computed on the graph of filaments*

We harness ideas and order parameters from graph theory (network theory) [19,26,27] to explore their ability to quantify the dynamics of actomyosin networks. The key strength of network theory is to assess the layout of complex networks [19,20,28] by incorporating far greater information about the high order of relations between nodes in the network. We reduce the complex actomyosin network to a graph representation by defining it into a coarse-grained network in terms of the spatial distance between actin filaments without direct reference to any ABP between them, as the local microscopic information of connectivity between ABPs and filaments is not always available experimentally. The graph representation $\mathbf{G} = (V, E)$ of a snapshot of an actomyosin network is defined a set of nodes $V$, which correspond to the filaments, and a set of edges $E$, which accounts for the relations between nodes. An edge exists between any two nodes if their two corresponding filaments in the system are within a cutoff distance $d_{\text{cut}}$. We define the distance $d(v, w)$ between filament $v$ and $w$ to be the minimum Euclidean distance among all pairs of their segments' centers of mass. Mathematically, the graph representation of the actomyosin network is defined as an adjacency matrix

$$\mathbf{A}_{vw} = \begin{cases} 1, & d(v,w) < d_{\text{cut}} \\ 0, & \text{otherwise} \end{cases}, \qquad (1)$$

where $d_{\text{cut}} = 200$ nm. The number of vertices in the graph $\mathbf{G} = (V, E)$ is $|V| = N_{\text{filaments}}$ and the number of edges $|E|$ varies depending on the spatial distribution of the filaments. Topological

measures of a graph can be used as order parameters at varying graininess from a "node", a "local", to a "community" viewpoint. They are sorted into three viewpoints: (a) the topology of pairs or triplets of nodes that quantify the node-level topology in a network, which is illustrated in **Fig. 2**(b) the local-level measures of centrality, which quantifies the connectivity or the 'importance' of local nodes in a network, and (c) the detection of community-level structures in a global topology of a network. The node-level measures of the networks are illustrated in **Fig. 2**. Below, we will explain other network-theory order parameters in greater details as they describe the mesoscopic features of a network (e.g. "local" or "community") more than node-level measures.

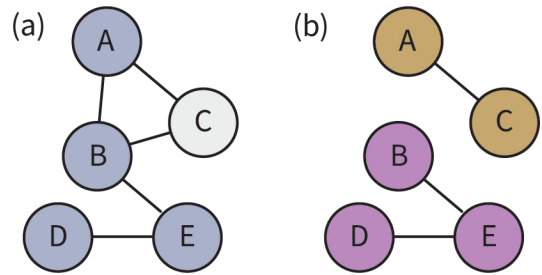

**FIG. 2**. Fundamental graph properties at a node level on two network models (a) and (b). Each model comprises five nodes, $V = \{A, B, C, D, E\}$. The difference between the network in (a) and the network in (b) is the set of edges connecting these nodes. Graph (a) has one triangle ($\{A, B, C\}$). While in graph (a) all nodes are connected, in graph (b) there are two separate components: $\{A, C\}$ and $\{B, D, E\}$. In (a), the clique number, which is the largest subset of nodes that are all directly connected, is 3, and the largest clique is $\{A, B, C\}$. In contrast, the largest clique in graph (b) is $\{A, C\}$. The degree of a node is the number of edges connected to it. For example, in (a), the degree of node B equals 3. The shortest path between two nodes is the minimal number of edges one needs to hop over to reach from one node to the other. In graph (a), the shortest path between node D and node A is D ↔ E ↔ B ↔ A. Nodes that are disconnected, as in the case of D and A in (b), are said to be separated by an infinite path, by definition. The betweenness of a node is the number of the shortest paths between any other nodes that go through this node. For example, in (a), the node B has the largest betweenness of 4, as the shortest path between D and A, between D and C, between E and A, and between E and C, all go through B.

Here, we list the order parameters and measures from network theory that we have found the most useful to characterize the complex morphologies of actomyosin networks in our study.

*(i) Graph Density and Mean Centrality:* Graph density measures the number of edges in a network relative to the maximum number of edges that could exist if all nodes were connected. It is defined as the ratio of the average node degree, $\bar{d} = \frac{1}{|V|} \sum_{u,w} \mathbf{A}_{uw} = \frac{2|E|}{|V|}$, over the largest possible node degree, $|V - 1|$:

$$d_{\mathbf{G}} = \frac{\bar{d}}{|V-1|} = \frac{2|E|}{|V| \cdot |V-1|}. \tag{2}$$

The degree of a node, $k_u = \sum_w \mathbf{A}_{uw}$, can also be used to define the "centrality" of a node, quantifying the importance of an individual node in a network (with the degree centrality defined by the number of edges a node has). A number of alternate definitions for centrality, which quantify the local topology of the network in specific ways, are described in the Supplementary Information. While graph density represents a normalized mean degree centrality of each node in the network, multiple order parameters based on the centrality of nodes, such as the betweenness and the eigenvector centralities, have been used.

*(ii) Assortativity:* In addition to considering solely node connectivity (through node degree), it is useful to quantify the local network topology in a complex network based on sorting the connections between nodes. The degree assortativity coefficient [29], as defined by Newman et al. [22,30], is an order parameter that explores the correlation between the nodes with similar degrees to be directly connected. The degree joint counting matrix, $\mathbf{J}_{lm}$, is the number of edges in a graph, linking a node of degree $l$ with another node of degree $m$:

$$\mathbf{J}_{lm} = |\{(u,v) : v, u \in E \wedge k_v = l \wedge k_u = m\}|, \tag{3}$$

where $k_u = \sum_w \mathbf{A}_{uw}$ is the degree of node $u$. The degree assortativity coefficient (often referred to as the assortativity) is defined as:

$$\rho = \frac{\sum_{l,m} l \cdot m \cdot (\mathbf{J}_{lm} - p_l p_m)}{\sigma_p^2}, \tag{4}$$

where $p_m = \sum_l \mathbf{J}_{lm} = \sum_l \mathbf{J}_{ml}$ (the probability of seeing a node with degree m), and $\sigma_p$ is the standard deviation of the set $\{p_m\}_{m=0}^{|V|-1}$. If there were no bias in the connectivity in a large network, we would expect $\mathbf{J}_{lm} = p_l p_m$, leading to $\rho = 0$ in a random network. We note that $\rho$ ranges from 1 to −1, with a positive (high) assortativity indicating that there is a central core of highly connected nodes and a negative (low) assortativity indicating a star-like topology in the network (with high-degree nodes connected to only low-degree nodes).

*(iii) Clustering coefficient:* In many physically and biologically relevant networks, it has been found that triplets of nodes share edges (form a `triangle', e.g. the triplet {A, B, C} in **Fig. 2**(a)) more often than would be expected by random chance [31]. This is referred to as a clustering coefficient in network theory literature[19], and is conceptually distinct from the clustering discussed in the gelation literature (see the Appendix). The normalized clustering coefficient, counting the number of triangles that exist in the network relative to the number that could exist, is computed using

$$\Gamma = \frac{1}{2|V|} \sum_{v \in V} \frac{\sum_{i,j} A_{vi} A_{ij} A_{vj}}{k_v(k_v - 1)}. \tag{5}$$

A high clustering coefficient implies that the connectivity is dense locally (with a propensity to develop triangles in the network), even if the graph density of the network is sparse. Note that in the actomyosin networks, a 'triangle' could imply to a physical triangle of filaments but given the coarse-grained nature of our graph could refer to many possible geometries. Of note, we often refer the clustering coefficient as the average clustering.

*E.2 Correlations in global connectivity*

In addition to the node-level (degree) and local-level (assortativity or clustering) order parameters, there are a variety of network measures that account for the global topology. These include communities and cliques of nodes, where a high- or complete-connectivity between nodes indicate regions of high connectivity within the graph. These network measures are conceptually distinct from the gelation representation of a "cluster", which accounts solely for the existence of a path between constituent filaments without regard for the density of connections between filaments in the cluster.

*(i) Community Structure*: Community structure is widely used to quantify global network topology [32], and indicates the presence of heterogeneity in the density of links in a network: if some nodes are more likely connected with each other than with the rest of the network, they may be said to form a community. In the network science literature [20], a "community" refers to a collection of nodes that are more highly interconnected between each other than they are with the rest of the network. This is distinct from the gelation formalism, where a presence or absence of a path between filaments determines the membership in a cluster of filaments (please see the Supplementary Information). The heuristic definition of a community has spawned an enormous literature on the subject of community structure in networks [32-35]. In this paper, we focus solely on a commonly used method: the Clauset-Newman-Moore greedy modularity maximization method [36]. The modularity $Q$ is a measure which evaluates the strength of division into distinct modules/communities in a network. We apply the commonly used greedy maximization technique to find the membership of the communities in a graph **G**. Denoting the community of a node $v \in$ V by $s_v$, the greedy modularity maximization algorithm assigns a unique community $s_v$ to each node $v \in$ V to maximize the modularity:

$$Q = \frac{1}{|E|} \sum_{(v,w) \in E} \left[ \mathbf{A}_{vw} - \frac{k_v k_w}{|E|} \right] \delta_{s_v, s_w}, \tag{6}$$

where $k_v$ is the degree of node $v$ and $\delta_{a,b}$ being the Kronecker delta function, which equals 1 if $a = b$ and 0 otherwise. This definition of modularity divides nodes between communities based on whether nodes within a community are more densely connected than would be expected by random chance. Using this definition of a community, we define the order parameter $N_{\text{com}}$ to be the number of communities in a graph **G**.

*(ii) Clique number*:  A clique in a graph refers to a subset of nodes that are fully connected with one another, and thus represent a more extreme heuristic definition of a community in a network than is realized using modularity maximization. We consider "clique number" the maximal clique size as an order parameter (schematically illustrated in **Fig.** 2), which is the largest subset of vertices from V such that each two nodes in the subset have an edge between them.

### F. Potential of Mean Force of Actomyosin Networks

In molecular dynamics simulations, the potential of mean force (PMF) is a common technique to investigate the dependency of the free energy on a specific order parameter. The PMF, or $\mathcal{U}$, is defined as the negative natural logarithm of the probability density function (PDF), or $P$, against some order parameter of a system [37]. If we consider a normalized order parameter $\xi \in [0, 1]$ and let $P(\xi)$ be its PDF, we can compute the PMF $\mathcal{U}(\xi)$, as elaborated further in the Supplemental Information Method section VI, using the relation:

$$\mathcal{U}(\xi) \stackrel{\text{def}}{=} -k_B T \ln P(\xi) . \tag{7}$$

The $\mathcal{U}(\xi)$ of an order parameter can be thought of as the free energy profile of the system with respect to $\xi$ along a trajectory, and the relation in Eq. (7) is applicable for both equilibrium and non-equilibrium systems at a constant temperature, $T$ [37,38].

## III. RESULTS

### A. Multilinkers with a high valency enrich arborization of actin bundles

We first show how multilinker's valency, $\nu$, changes the topology of actomyosin networks at a high motor content with time by examining the snapshots from the Cyosim simulations for $\nu = 2$, 3, and 6 in **Fig.** 3. Initially, a time $t = t_0$, all three networks start out as dispersed filament network (we called it the "solution state") and form small lumps at $t = 50$ s, as seen in **Fig. 3**(a). Then they form large fuzzy assemblies (we called it "disordered gel") by the time $t = 100$ s. The system with hexavalent ($\nu = 6$) multilinkers has the most filaments participating in the formation of the largest assembly. At $t = 200$ s, in **Fig. 3**(b) we observed the formation of ordered bundles grow thicker as time evolves, and, most noticeably for multilinkers of high valency ($\nu = 6$), the thick bundle breakouts into a tree-like structure as seen at both $t = 400$ s and $t = 600$ s in **Fig. 3**(b). This is a new state of ordered bundles forming a large stalk with one or more thinner bundles adjoining the main stalk. We refer to this topology of dense bundles attached to thin branches as the "arborization" of the actin network.

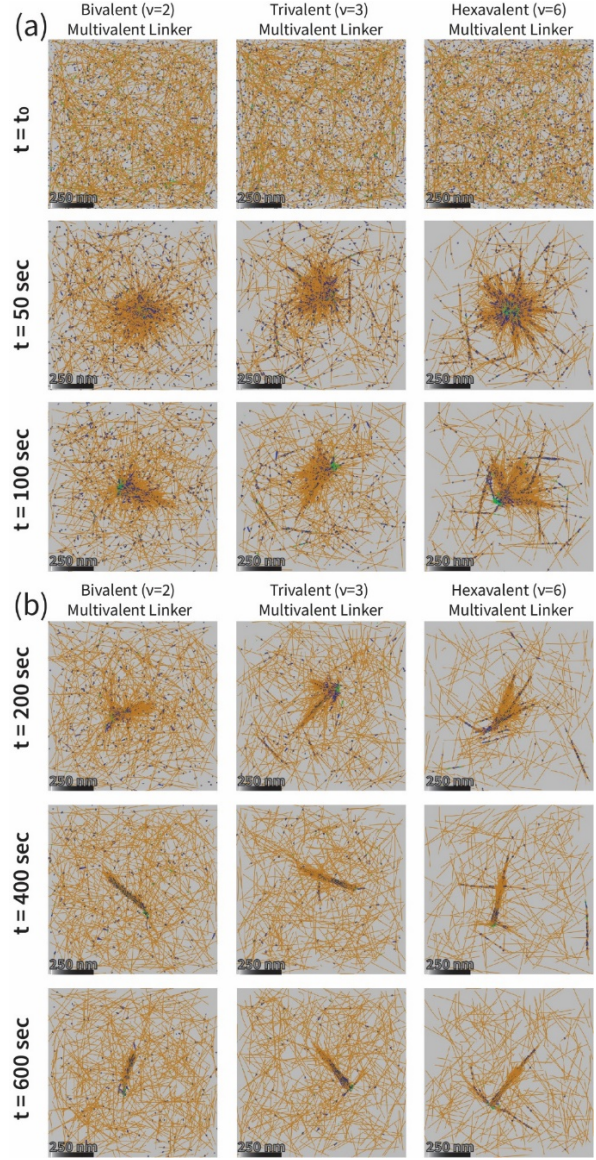

**FIG. 3**. Snapshots from Cytosim simulations at three multilinker valency conditions. All simulations are in a $1\mu m^3$ box with the same components of filaments, motors, multilinkers, and α-actinin like crosslinkers. These three systems differ only by the valency of the multilinkers. On the leftmost column in (a) and (b), the multilinkers are bivalent ($\nu = 2$). On the middle column, the multilinkers are trivalent ($\nu = 3$). On the rightmost column, the multilinkers are hexavalent ($\nu = 6$). The time evolves from the top panels to the bottom panels. In (a) the snapshots are in an increment of 50 seconds, starting at time $t = t_0$, where $t_0 = 1$ sec. In (b) the snapshots are in an increment of 200 seconds, ending at $t = 600$ sec. Noticeably, for a system with hexavalent multilinker at $t = 400$ sec and $t = 600$ sec in (b), the main stalk in the middle of the snapshot is a thick bundle with adjoint thinner bundles (a phenomenon we called arborization).

## B. Graph representations of complex actomyosin networks

Regardless of whether these active networks of filaments thicken the ordered bundles or arborize into thinner bundles from a stalk, an immediate challenge is to identify proper descriptors that capture the complexity in the evolving actomyosin network as the ABPs remodel the filaments in space. Our strategy is to mirror a snapshot of actomyosin network into its graph representation and apply tools from network theory to capture the topological properties of the underline actomyosin network. **Fig. 4** shows four such graph representations of snapshots from **Fig. 3** representing the simulations for $v = 2$ and $v = 6$ at $t = 1$ s (**Fig. 4**(a) and **4**(b)) and at $t = 600$ s (**Fig. 4**(c) and **4**(d)).

The topologies of these two networks appear distinctive after 600 seconds of temporal evolution. These differing features of these networks have been clearly captured by the order parameters from network theory in **Fig 4**. With a greater graph density of 0.42, the actomyosin networks containing the hexavalent multilinker with $v = 6$ creates a centralized, higher interconnected nodes than the networks containing the bivalents multilinker (a graph density of 0.14). The former has an average node degree of 422.2, whereas the latter has an average node degree of 140.7.

By measuring the betweenness centrality of each node (defined in the Supplementary Information), we quantify the importance of this node in connecting any two other nodes by being in the "shortest path" in-between them. We visualize this information of betweenness by coloring them in a scaled purple color in **Fig. 4**. There are far more dark purple nodes after 600 seconds of simulations in **Fig. 4**(d) for $v = 6$ than **Fig. 4**(c) for $v = 2$, indicating that there are more nodes serving in the "shortest paths" between any two nodes within the actomyosin systems containing the hexavalent multilinkers than those containing bivalent multilinkers. Alternatively, this observation is also supported by the fact that the average path length between any two nodes is 3.1 in the system with hexavalent multilinkers whereas the average path length becomes 6.1 in the system with bivalent multilinkers.

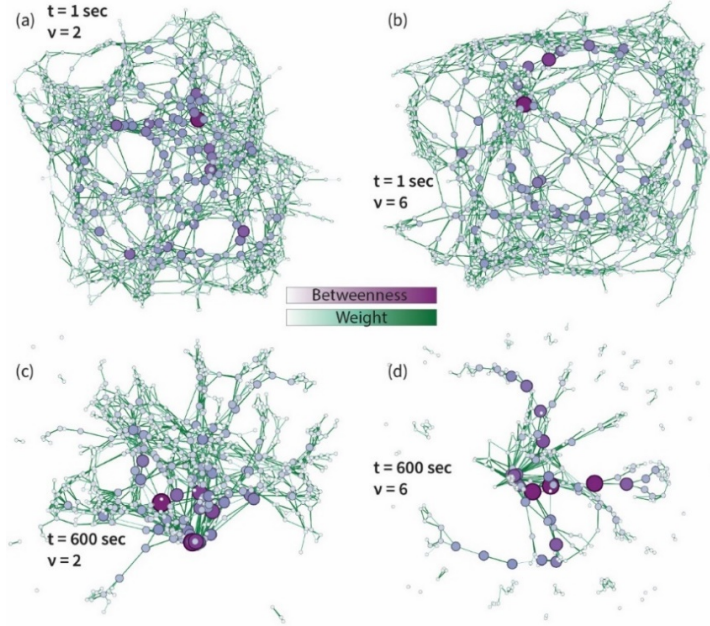

**FIG. 4**. Graph representations of four actomyosin network snapshots shown in **Fig. 2**. The graph visualization of a system with bivalent ($v = 2$) multilinkers (a) at $t = 1$ sec and (c) at $t = 600$ sec. The graph visualization of a system with hexavalent ($v = 6$) multilinkers (b) at $t = 1$ second and (d) at $t = 600$ sec. We produced these graphs by iterating the same visualization procedure starting from a random network layout produced by OpenOrd [39] and followed by Yifan Hu's visualization algorithm [40], both with the default parameters in Gephi 0.92 [41]. The betweenness of a node counts the number of times it lies on the shortest path between any two other nodes (please see the Supplementary Information).

### C. Correlation of the order parameters from polymer physics, gelation theory, and network theory

The rich actomyosin structures shown in **Fig. 3** have prompted us to seek appropriate order parameters to characterize the complexity and heterogeneity of actomyosin dynamics in space over time. We have compared order parameters inspired by three domains of theories: polymer physics, gelation theory, and network theory, as described in the Methods section and Supplementary Information. **Fig. 5** shows the Pearson correlation of all these order parameters evaluated for each actomyosin configuration from the simulations performed using Cytosim. Each category of order parameters presents a different insight into the actomyosin morphology. The three order parameters from polymer physics (defined in the Appendix) are highly correlated. When we examine their time signals it turns out all the three fluctuate around a global mean value, as illustrated in the cumulative distribution functions in **Fig. S1** and **Fig. S3**(g). Even though the polymeric order parameters are insufficient to describe the morphology of the network, their power spectral density (PSD) [42] profiles show that high motor content increase the power density by two orders of magnitude (20 dB), as shown in **Fig. S2**. The insufficiency of the polymeric order parameters to describe the dynamics of our complex systems contrasts with previous computational studies at which the radius of gyration, $R_g$ and the shape parameter, S (both introduced in the Appendix) were appropriate order parameters to describe actomyosin networks containing Arp2/3 (actin-related protein 2/3) which nucleates on a mother filament and grow a daughter branch at a fixed angle [13,43]. However, these polymer physics order parameters are no longer sufficient in addressing the increased complexity in a network topology when multilinkers are present and the number of filaments in the system increased significantly from 50 to 1,000. Notably, all three polymer order parameters are correlated within their own category, but they are nearly uncorrelated with the order parameters from gelation theory or graph theory (**Fig. 5**). We showed that order parameters from gelation and network theories are suitable for assessing the complex and diverse morphologies of the simulated actomyosin networks by showing their positive correlations in **Fig. 5**, even though the ways to define these order parameters are different. The gelation-theory order parameters are computed from a list of connections between filaments directly made by ABPs, whereas the network-theory order parameters are computed from the distances between filaments that are not necessarily in direct contacts. Noticeably in **Fig. 5**, most gelation-theory order parameters are clustered together by themselves, and most network-theory order parameters are clustered together by themselves. Yet, there is still positive correlation between the two categories of order parameters. (For further analysis using gelation theory order parameters, please see the Supplementary Information). Finally, there are two order parameters that parameters that negatively correlate with the rest. One is the node degree assortativity (or simply assortativity) that measures whether the filaments with the same node degree prefer to connect or not; the second is the number of gelated clusters ("Number of Clusters" in **Fig. 5** which is defined in the gelation theory in the Appendix). The assortativity has a negative correlation with most order parameters, because, as we discussed later, the higher the valency of multilinkers the lower the assortativity in the networks. The total number of gelated clusters is negatively correlated, because as time evolves the networks in our simulations prefer to connect more filaments together into a large ordered gel.

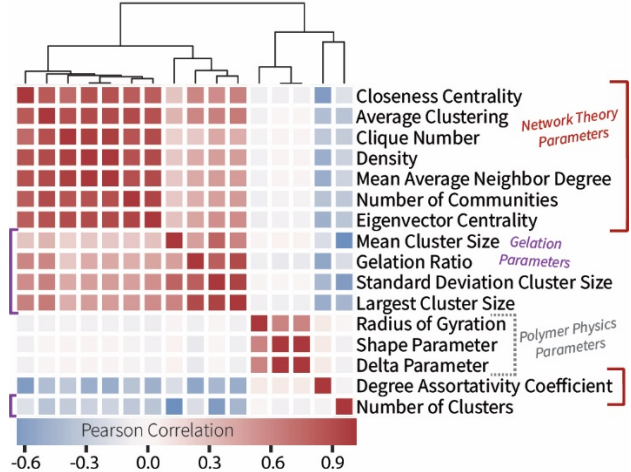

**FIG. 5**. The Pearson correlation between the order parameters used for this study colored in a heat map. The order parameters are sorted by the pairwise correlation distance. To aid visualization, they are further grouped into three categories by their physical order parameter: polymer physics (labeled in grey dashed brackets in the right), gelation theory (labeled in purple brackets in the left), or network theory (labeled in red brackets in the right).

### E. Analysis of actin filament networks in graphical representation

Although the order parameters from gelation theory captures the sizes and the distribution of filaments that are linked with ABPs (please see the Supplementary Information), it is less sensitive to the formation of complex structures beyond immediately connected actin filaments. Therefore, they are unable to capture higher-order organization in a complex network. We characterize the higher-order complexity in a network by applying the network-theory order parameters. To achieve that, we first converted actomyosin networks into graphical representations where the heterogeneously distributed filaments are nodes (see section **B** about graph representations). We then apply appropriate descriptors from a wide range of network-theory order parameters with varying graininess to characterize their patterns in connectivity and sizes, and to detect communities.

We measure how interconnected the network is with the average clustering value that ranges from 0 to 1. This measure is similar to asking a question, "do my neighbors know each other or not". In **Fig. 6**(a) at low motor content, multilinkers with higher valency ($v \geq 3$) gradually create a more interconnected network than those with $v = 2$ by showing increased clustering coefficients over time. At high motor content (**Fig. 6**(b)), almost all multilinkers create highly interconnected networks with clustering coefficient above 0.6 . Such measures show that multivalency produces distinct topologies in networks with either high or low motor content. This knowledge is absent from gelation-theory order parameters.

At low motor content, there is less variation in the network topologies as the time signals of the clique number in **Fig. 6**(c) suggest. Nonetheless, in **Fig. 6**(d) at high motor content systems with multilinkers of higher valencies create larger cliques in the network. Namely, although multilinkers increase the number of possible binding sites to actin in the actomyosin networks, valency plays a small role in determining the higher-order assembly of the network when motor content is low. On the other hand, at high motor content, motors further increase the higher-order organization of the network by enhancing the local assembling of filaments in the graph, as shown by a larger clique number in **Fig. 6**(d). The motor activity allows multilinkers of higher valencies to amplify the clique number of the network by efficiently bundling multiple actin filaments at once.

As the systems become more complex and heterogeneous, we asked whether these ordered bundles further collapse into thicker bundles, or they arborize into thinner branches. We are essentially asking "to which community do ordered bundles belong?". We applied modularity maximization to count the number of distinct communities [36]. We show that as the number of communities increases, so does the hierarchy in a network, particularly at high motor content and with multilinkers of high valency. Interestingly, the profiles of the number of communities in **Fig. 6**(e)-(f) resemble those of the average clustering profiles in **Fig. 6**(a)-(b). It exemplifies that the roles of active motors and passive multilinkers in increasing the complexity of a graph by increasing the number of distinctive communities, instead of collapsing the actomyosin networks into a single homogenous community (or a gelation phase).

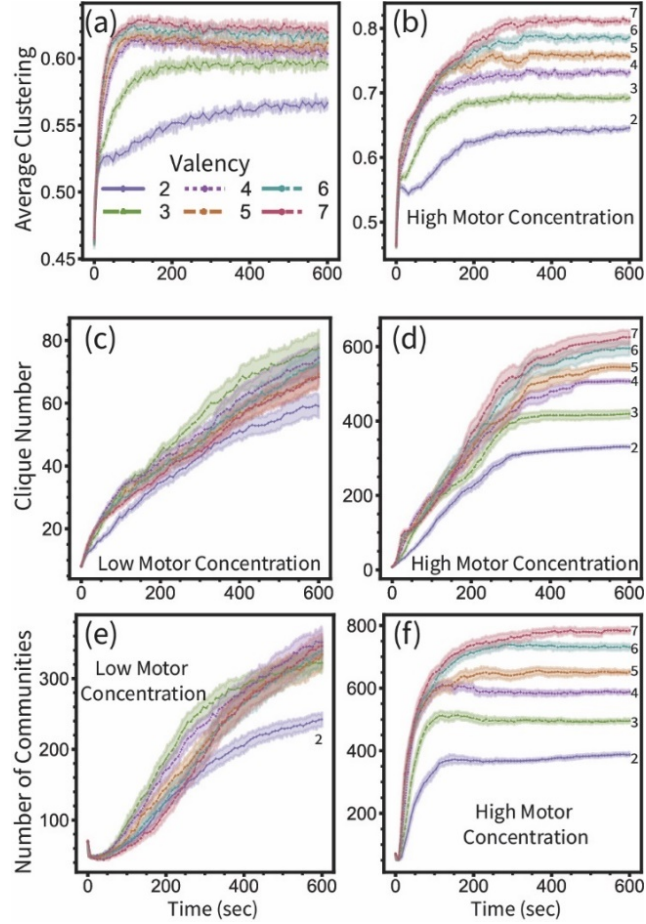

**FIG. 6**. Time signals of the network-theory order parameters computed from the simulations of actomyosin networks with multilinkers. All the systems made of the same compositions (filaments, crosslinkers, motors, multilinkers) with multilinkers at various valencies and at low and high motor content. The right column (a, c, e) shows the time signals from systems with low motor content and the right column (b, d, f) show the time signals at high motor content. Three order parameters are shown: (a, b) the average clustering defined in Eq. (5), (c, d) the clique number of the network, and (e, f) the number of communities [36].

## F. Decreases in node degree correlation (assortativity) suggest arborization

Next, we zoom out to a nonlocal level to view the hierarchical ordering of nodes (filaments) in a network by examining the node degree assortativity. Assortativity ranges between $-1$ and $1$. A positive assortativity indicates that nodes with similar degrees are more frequently connected to each other. In the graphical representations of the actomyosin networks in our simulations, assortativity goes above $0.35$ over time in **Fig. 7**(a,c). Assortativity of systems with higher valency multilinkers attenuates over time compared to those with lower valency multilinkers. At low motor content, assortativity plateaus (**Fig. 7**(a)) as filaments groups into thick bundles (**Fig. 7**(b)). At high motor content, the profiles first sharply drop and rise again, instead. The extent of recovery depends on valency. The profiles for systems with $v > 3$ hardly rise up and the assortativity remains low (**Fig. 7**(c)). It is because filaments first form thick bundles and then arborize to form tree-like branches (**Fig. 7**(d)) in the presence of high valent multilinker and high content of motors. In other words, actomyosin networks actively branch out rather than forming into a single stalk.

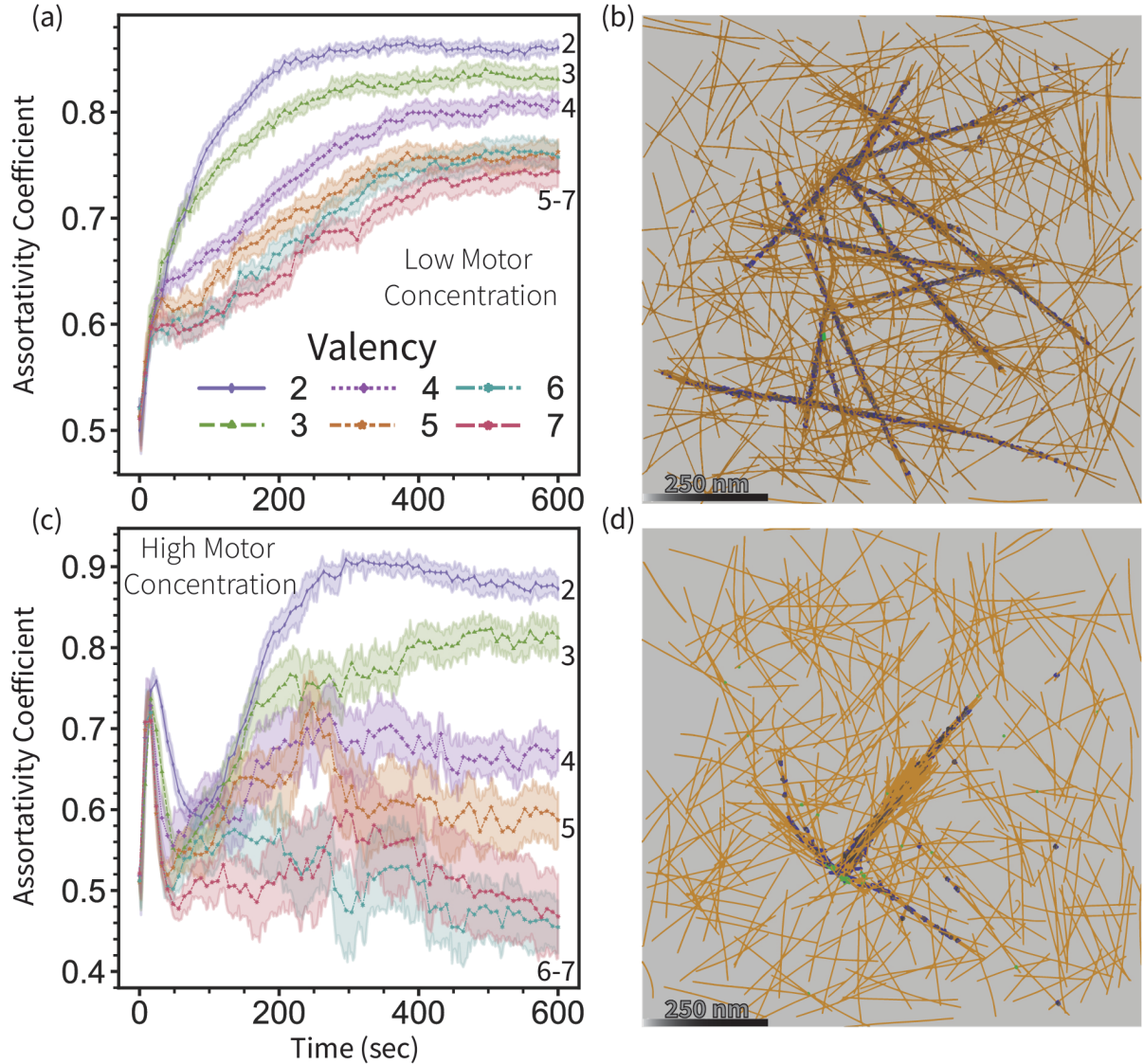

**FIG. 7**. The temporal evolution of the node degree assortativity coefficient (Eq. (4)) of actomyosin networks in the presence multilinkers with varying valency. The top row is from simulations at low motor content. The bottom row is from simulations at high motor content. The snapshots in (b) and (d) are taken from the simulations at $t = 600s$ where the multilinker valency $v = 6$. The actin filament bundle branches out at a high concentration of motors, while actin filaments form a mesh grid of think bundles in space at a low concentration of motors.

## G. Network theory order parameters reveal higher order assemblies in actomyosin network

To reveal the spatial heterogeneity of actomyosin networks at a community level, we applied the graph density order parameter that measures how "dense" (non-sparse) the adjacency matrix of a graph is. In **Fig. 8**(a), the graph density shows that even for low motor activity, the presence of multilinkers with $v > 2$ increases the density by 2-fold, although the overall graph density still remains low.

Notably, in systems with high motor content in **Fig. 8**(b), the graph density is 10-fold larger than system with low motor content. Not only they become denser, the profiles of graph density for high motor content becomes more complex and resemble logistic curves in **Fig. 8**(b). For example, the profile for trivalent multilinker ($v = 3$) evolves in steps characteristic of a double logistic pattern: the graph density evolves in the first 100 s and remains around density of 0.1 until $t = 200$ s. Next, it rises to 0.2 and plateaus after $t = 300$ s. For pentavalent multilinkers ($v = 5$), the graph density profile even undergoes more than two transitions and the graph density increases by another 50% to 0.34 compared to those for $v = 3$.

This observation of three transitions suggests that there are at least four phases of organization. As shown in the simulation snapshots in **Fig. 3**, the first three phases from the simulations for trivalent multilinkers ($v = 3$) are the "solution" phase of the actomyosin at $t = t_0$ in **Fig. 3**(a); the second phase of the network is a gel at $t = 200$ s in **Fig. 3**(b); and the third phase is the order bundled phase of actomyosin at $t = 600$ s. To visualize the fourth phase of "arborization", we turn to the last snapshot for hexavalent multilinkers ($v = 6$) at $t = 600$ s. The descriptors from network theory are able to capture the details of these transitions better than those from polymer theory or gelation theory.

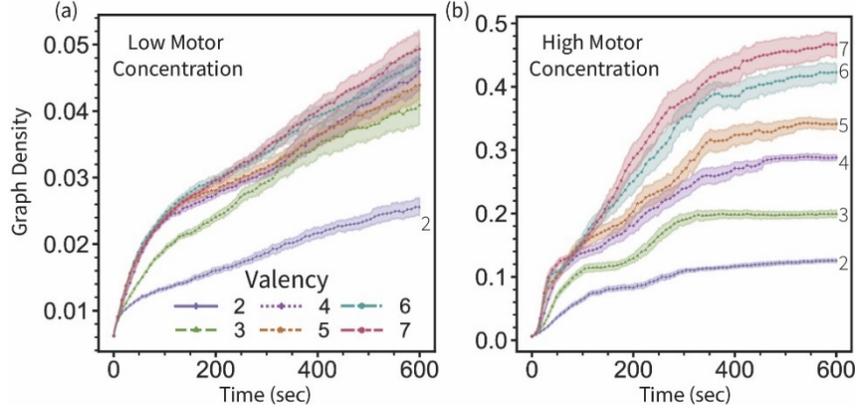

**FIG. 8**. Graph density of actomyosin networks in the presence of multilinkers with varying valency. The graph density is plotted (a) in a low motor content and (b) in a high motor content. Each profile in (a) and (b) is averaged of 30 independent trajectories with the same species parameters with a confidence interval of CI = 90%.

To further investigate the utility of using network theory order parameters, we compare the profiles of the potential of the mean force (PMF) established with a network-theory order parameter and with a gelation-theory order parameter. We computed the PMF against the average neighbor degree of nodes from network theory in **Fig. 9**(a) against the sizes of gel clusters from gelation theory in **Fig. 9**(b). For $v = 6$, in addition to the two global minima near $\xi = 0.0$ and $\xi = 0.7$, there are two intermediate local minima around $\xi = 0.2$ and $\xi = 0.4$ in **Fig. 9**(a). In contrast, only one intermediate state is apparent on the PMF characterized by the gelation-theory order parameters in **Fig. 9**(b). It is evident that using only order parameters from gelation theory is insufficient to capture the complexity in a dendritic growth from actomyosin networks, since the gelation formalism is not sensitive to the growing of multiple thinned branches out of a common stalk. Therefore, it is essential to employ order parameters from network theory as descriptors to capture these higher-order, hierarchical topologies.

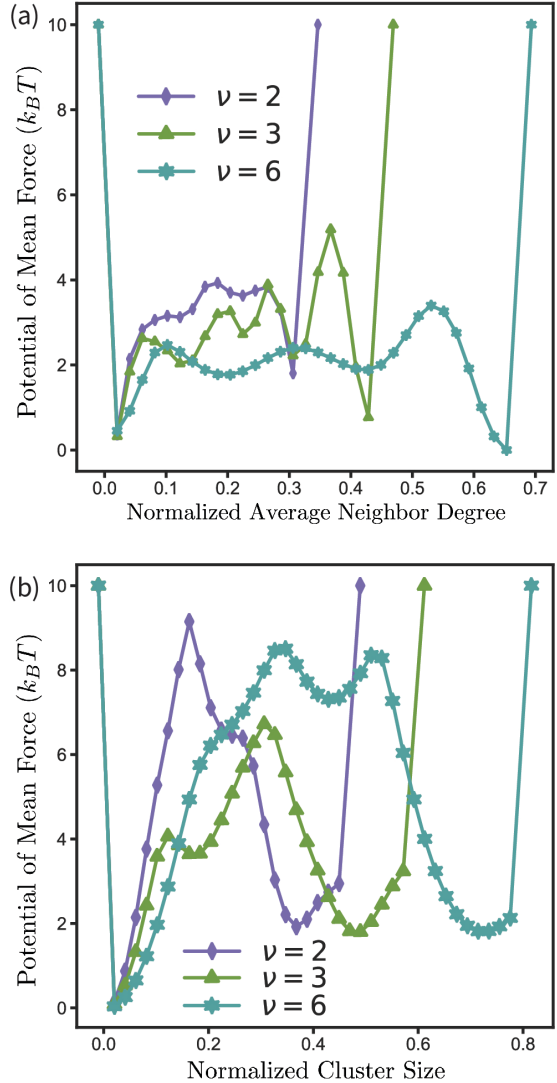

**FIG. 9**. Potential of the mean force (PMF) curves of actomyosin systems with three types of multilinkers at high motor content. (a) The PMF is plotted against the average neighbor degree from network theory. (b) The PMF is plotted against the cluster size from gelation theory.

### H. Network-theory order parameters reveal a rich state diagram of actomyosin networks for experimental prediction

We have identified network-theory order parameters necessary for describing local or non-local features in a complex network. We further set up our hypothesis that in addition to multilinkers' valency, the motor content also plays a significant role in ordering the hierarchy of a complex network. We first established a state diagram by compiling these network-theory order parameters from all of the steady-state simulations (the last 20 seconds of any trajectory) in plane of both the motor content and the multilinker content. State diagrams in **Fig. 10** reveal the states of solution, gelation, ordered bundles, and arborization, and provide readily applicable guidelines for experimentalists to probe these phases and transitions in reconstituted or living samples.

When reviewing the state diagram of graph density in **Fig. 10**(a) we recognized that the graph density increases with the multilinker content as well as the motor content in a non-monotonic fashion. This plot suggests that the actomyosin networks become densest when the motor-filament as well as the multilinker-filament ratios are about 2: 1, and the graph density diminishes at the highest motor or at the highest multilinker content. This feature is interesting because it signifies higher-order assemblies of actomyosin networks beyond a homogenous gelation state.

We exemplify the formation of higher-order assemblies by scanning the 2D state diagrams at a constant multilinker to filament ratios (the horizontal gray dashed lines in **Fig. 10**(a)-(b)). We noted that both graph density (**Fig. 10**(a)) as well as degree assortativity (**Fig. 10**(b)) decrease at high motor content. When assortativity in a network decreases, mathematically we interpreted that there are less nodes being connected to another node with a similar node degree. How does it relate to changes in the morphology of actomyosin network? We observed from the simulations that it comes from a distinctive feature where an ordered bundled stalk thins out into several dendritic branches, rather than growing into a common stalk. Such a dendritic process also reflects in decreased graph density from 0.3 to 0.06 in **Fig. 10**(a). The state diagram captures the emergence of a new "arborization phase" as shown in **Fig. 7**(d).

Under certain conditions, filaments form bundles without higher-order organization of thick stalks or arborized morphologies. When we set up the system with a low motor content on the state diagram (the vertical black dashed lines in **Fig. 10**(a)-(b)), the graph density does not vary (**Fig. 10**(a)) while assortativity does. Assortativity first increases from 0.45 to 0.65 at low content of multilinkers and then decreases from 0.65 to 0.45 at high content of multilinkers (**Fig. 10**(b)). This signifies that multilinkers facilitate the formation of evenly distributed ordered bundles (as shown in **Fig. 7**(b)) in a network, without self-assembling into a higher-order scaffold such as a common stalk with dendritic branches.

Another order parameter that possibly reveals a rich variety of local topology is the clique number, shown in **Fig. 10**(c). We see that the clique number remains low in systems with multilinker content of low valency such as $\nu \leq 3$ (vertical dashed line at $\nu = 3$), while it increases quickly for high valency such as $\nu > 3$ (e.g., along the vertical dashed line at $\nu = 6$). This signifies how multivalency increases the complexity in connectivity among nodes in the network.

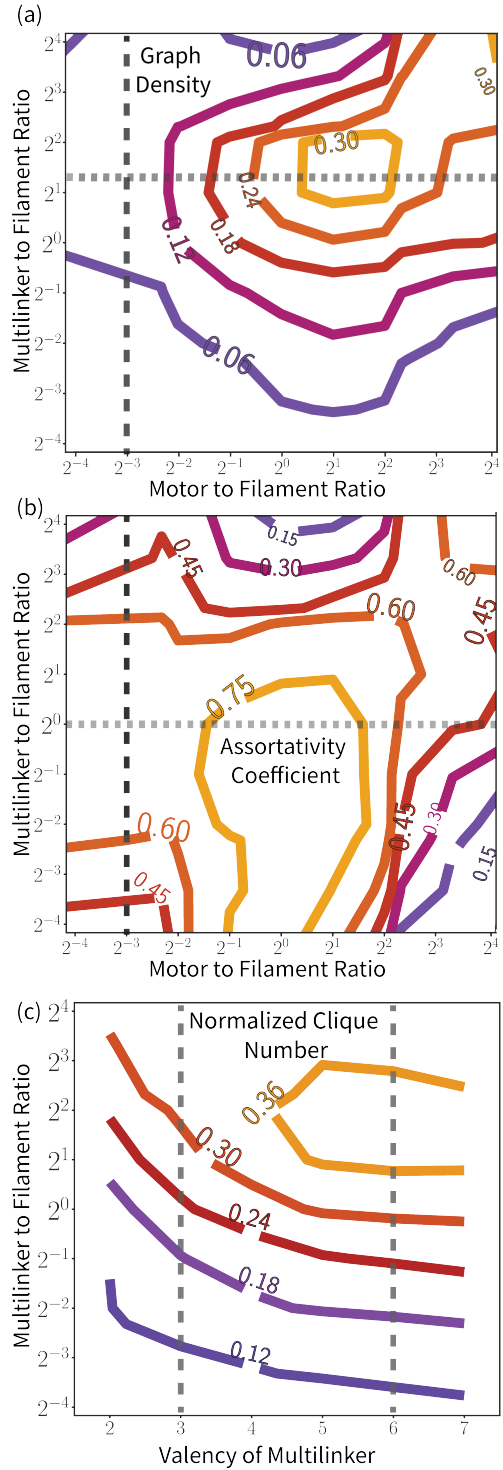

**FIG. 10**. State diagrams of network-theory order parameters. The state diagram of (a) graph density and (b) assortativity in plane of multilinker content and motor content. (c) The state diagram of clique number normalized by the number of nodes (filaments) in the system in plane of multilinker content and valency of multilinker. For each data point in the state diagrams, the information from the last 20 seconds in a steady state from each trajectory was taken for averages. Of note, to create a continuous plot in (c) along the valency of multilinker axis, a Gaussian filter was used to smooth the signal along the x-axis.

## IV. DISCUSSION

### A. Network theory approach is necessary to characterize the dendritic features on actomyosin networks

We have simulated the evolution of actomyosin networks mediated by actin-binding proteins using Cytosim. The simulated actomyosin networks are complex as the filaments form higher-order architectures under the influence of passive linkers and the active motors. As the system becomes increasingly heterogenous, we found that order parameters from polymer physics and gelation theory lack the ability to characterize the higher-order assemblies. By converting the actomyosin network into a graph representation with nodes and edges, we are able to quantify the hierarchy of a dendritic network with network-theory order parameters that provide principles for combining multiple local models into a joint global model.

The ability to compute global and local properties from the same graph allows us to connect local and global levels of information. By combining these descriptions that connect microscopic properties to macroscopic phenomena, we reveal the principle of self-organization that encompasses multiple length and time scales, allowing experimental designs for validation. In the dynamics of the global graph density in **Fig. 8**, we noticed a subtle double sigmoidal shape suggesting that there is a two-step transition in high motor content, signifying transitions from solution, gelation, and ordered bundles. When the valency of multilinkers increases to 6 or higher, there are multiple-step transitions that relate to higher-order dendritic structures. Global parameters such as clique numbers in **Fig. 10**(c) allow us to scrutinize whether local properties such as valency of multilinkers can potentially impact global morphology. Such detailed descriptors to observe complex properties of an active matter are not captured by using either polymer or gelation order parameters.

By mapping an actomyosin network into a graph, our work shows that many interesting features of actomyosin networks can be captured from the connectivity between filaments created by actin-binding proteins, or alternatively by considering only the distances between filaments without explicitly considering actin-binding proteins. Because a matrix of connectivity is a reduced representation of the system, this approach with the descriptors (order parameters) from network theory reflects faithfully on the reorganization of actomyosin networks mediated by motor and actin-binding proteins. Our approach will not only be useful to analyzing actomyosin networks from computer simulations, but also to analyzing their images from high-resolution experiments where the clusters of the filaments are measured, but not necessarily the positions of ABPs. One could use adaptive rheology, as demonstrated by Gupta et al. [44], and record the dynamics of the F-actin via their fluorescent labeling using RFP-Ftractin [45]. To compute the graph representation from a rheology image we suggest to use classical image processing techniques such as Canny edge detection algorithm [46] to identify filament segments. From those segments one could construct a graph representation, which could be used, similar to how we constructed the graphs from Cytosim.

## B. Multilinkers' valency and motor proteins drive the complexity of actomyosin networks

Using the network-theory order parameters provided new insight into the hierarchy of self-assembly that forms actomyosin networks. We observed that such details cannot be adequately characterized by a two-state formalism (solution and gel). In addition to these two states, we observed another two states, ordered bundles and arborization (dendritic bundles), depending on the conditions of multilinker valency, multilinker concentration, and motor concentration. We therefore use them as variables and created three state diagrams in **Fig. 10**, illustrating how they contribute to the complex morphologies of actomyosins. The concept of creating these state diagrams follow the experiments and data analytics from the protocol by Bendix et al. [12]. They explored the role of α-actinin and motors on actomyosin architectures by varying their content with respect to actin monomers. In one dimension, by varying the motor content, they were able to explore how its active process contributes to the contraction of the actomyosin networks. In the other dimension, by varying the linker content, they were able to identify an agonistic mechanism that oppose contractions. Their work has inspired development of computational approaches [13,43] at a mesoscopic scales (including ours).

Our actomyosin networks are more complex than Bendix's, because here we have included the valency of multilinkers as another key variable that drives the complexity of the morphology. While motor proteins actively contract the actin filaments by exerting forces on them fueled by ATP hydrolysis, multilinkers such as CaMKII are able to bind more than two filaments at once. Multilinkers with high valency provide more ways to bundle filaments than bivalent crosslinkers, such as α-actinin. Using network-theory order parameters such as graph density or assortativity to describe global features, we are able to make predictions on the conditions that drive higher-order topologies of actomyosin networks such as ordered bundled state or a thick stalk with arborized, dendritic branches. Also using network-theory order parameters such as the clique number of a graph representation, we are able to establish hypotheses on the role of multilinkers' valency in increasing the complexity of a network by growing the local connectivity among individual nodes (i.e., filaments).

## C. Actomyosin structures mediated by multivalent actin-binding proteins may shape neural dendritic spines

Our work was inspired by the unique feature of multivalency in CaMKII that is important to the formation of dendritic filaments and actin bundles in dendritic spines of a neural cell[7,13]. (Please see Supplementary Information for the Relevant Background Knowledge on CaMKII and dendritic spines). CaMKII has a dual role – one role is to amplify the calmodulin-mediated calcium signaling by auto-phosphorylating all the CaMKII units; the other role is to reorganize the actin filament network and to maintain the stability of actin filaments [16]. In our recent work we have solved the seemingly conflicting role by understanding the molecular assembly of Actin/CaMKII complex [8]. Although high CaMKII content promotes highly stable actin bundles that last for days [16], once calcium-bound calmodulin (CaM) activates CaMKII in a CaMKII-bound actin bundles, CaMKII quickly releases actin in less than a second [5,6]. Another prominent feature of CaMKII that stands out from other bivalent ABPs is its high valency – binding multiple actins at once or form geometrically complexes with actin filaments [7,8]. With this distinctive feature, we speculate it plays a significant role in driving the morphology of a dendritic spine, from a filipodia conformation[47], the neck of a spine, to its mushroom-like head structure.

With the simulations we executed with Cytosim and the network-order parameters, we revealed the impact of multilinkers' valency in driving the complexity in the dynamics and properties of actomyosin networks. With global order parameters such as graph density as well as assortativity, we show that high multilinker content indeed create ordered bundles. Whether it spawns more ordered bundles or arborizes into higher-order, dendritic thin branches will depend on the motor content. At a low motor content, filaments grow into ordered bundles. At high motor content, it generates dendritic actin structures, arborizing thin branches from a thick stalk. With these useful order parameters from network theory, we have provided state diagrams that could be validated experimentally.

### E. Concluding Remarks

Our results highlight the relevance of network-theory order parameters for studying the complex structure and dynamics of actomyosin networks by projecting an actomyosin network on a reduced representation of a graph. Leveraging network theory, we characterize the higher-order topologies of actomyosin networks. We discovered higher-order complexes such as ordered bundled states and the arborization states, and established state diagrams in plane of motor and linker contents that enable experimental validations. We believe that the representation of actomyosin networks as graphs would be useful beyond the scope of our analysis. It may help to automate the process of coarse-graining using graph neural network [48] or it could lead to the integration of nonequilibrium physics theories [49-51] with statistical physics on probabilistic graphical models [52], that further our understanding of the principles of molecular self-assemblies that lead to emergent biological functions.

### Appendix

Herein, we define the order parameters used from polymer physics and gelation theory. The three polymeric order parameters are derived from the moment of inertia tensor defined from the distribution of filament vertices in space. The gelation order parameters are defined on filaments and how they are connected with one another by ABPs.

*1. Distribution of actin vertices in space*

We explored the dynamical changes in shape and structure of networks in the presence of multilinkers, for both low and high motor concentrations. We used three order parameters from the protein folding field [53]: the radius of gyration $R_g$ (the variance of all the positions of filament vertices), which is a proxy for the macroscopic structure, and the asphericity and shape parameters [13] that quantify the spatial asymmetry in the network. In order to define these order parameters, we first define the moment of inertia tensor of filament vertices at time $t$:

$$\mathbf{T}_{\alpha\beta}(t) = \frac{1}{2N^2}\sum_{i,j=1}^{N} \left(\mathbf{r}_{i\alpha}(t) - \mathbf{r}_{j\alpha}(t)\right)\left(\mathbf{r}_{i\beta}(t) - \mathbf{r}_{j\beta}(t)\right), \tag{A.1}$$

where $\mathbf{r}_{i\alpha}(t)$ is the $\alpha$-component of a filament vertex, $N$ is the number of filament vertices, and $\alpha, \beta \in \{x, y, z\}$ are the indices of the Cartesian elements. Then the radius of gyration is given by the sum of the eigenvalues, $\lambda_1 \geq \lambda_2 \geq \lambda_3$ of $\mathbf{T}$ at time $t$:

$$R_g(t) = \sqrt{tr\mathbf{T}(t)} = \sqrt{\sum_{i=1}^{3} \lambda_i(t)}. \tag{A.2}$$

We let $\bar{\lambda}(t) = \frac{tr\mathbf{T}(t)}{3}$ be the average tensor trace of $\mathbf{T}(t)$ and define [53,54] the asphericity,

$$\Delta(t) = \frac{3}{2}\frac{\left(\sum_{i=1}^{3}\left(\lambda_i(t)-\bar{\lambda}(t)\right)^2\right)}{(tr\mathbf{T}(t))^2}, \tag{A.3}$$

and the shape parameter,

$$S(t) = \frac{\prod_{i=1}^{3}(\lambda_i(t)-\bar{\lambda}(t))}{\left(\frac{1}{3}tr\mathbf{T}(t)\right)^3}. \tag{A.4}$$

The shape order parameter [53] specifies how prolate ($S > 0$) or oblate ($S < 0$) is the conformation of an actomyosin network, while asphericity measures how an actomyosin network conformation differs from a perfect sphere ($\Delta = 0$), and occur over the range $-\frac{1}{4} \leq S \leq 2$ and $0 \leq \Delta \leq 1$. These order parameters quantify the physical size and shape of the system as a whole, without accounting for the connectivity between any filaments due to the presence of crosslinkers or motors.

## 2. Gelation of actin filaments into clusters governed by actin-binding proteins

Order parameters that explicitly account for the connectivity give a complementary approach to quantify the dynamics of actomyosin networks. Previous work [13] has explored the sol-gel phase transition with respect to the molecular connectivity of ABPs and filament, which suggests a number of potential order parameters based on the size of filamentous clusters. The $i$-th cluster at time $t$ is a gelated group of filaments connected by either motors, linkers, or both; we denote the number of filaments in the $i$-th cluster as $N_{c_i}(t)$. By convention, a pair of filaments forms the smallest cluster of size two, hence, $N_{c_i}(t) > 1$ always holds. We denote $N_c(t)$, as the total number of clusters in the system at time $t$, and denote $N_f$ as the number of filaments in the system. The mean gelation ratio (the fraction of filaments found contained in any cluster),

$$\bar{N}_{\text{gel}}(t) = \frac{1}{N_f} \sum_{i=1}^{N_c(t)} N_{c_i}(t), \tag{A.5}$$

is a natural choice for an order parameter, as it quantifies the number filaments that are linked to other filaments. Complementary order parameters include the normalized mean cluster size

$$\bar{\mu}_c(t) = \frac{1}{N_f} \sum_{i=1}^{N_c(t)} \frac{1}{N_c(t)} N_{c_i}(t), \tag{A.6}$$

and the normalized largest cluster size

$$\bar{N}_{\max}(t) = \frac{1}{N_f} \max_{1 \leq i \leq N_c(t)} \{N_{c_i}(t)\}, \tag{A.7}$$

which measure the typical size of a cluster or the size of the largest cluster. The reasoning behind normalizing gelation-related quantities by the number of filaments is to map them onto the segment [0,1], with 0 indicating a completely unclustered network, and 1 indicating a network with a single large cluster. Eqs (6-8) quantify the degree of crosslinking within a network in similar ways, and are expected to be correlated. The variance of cluster sizes within the network,

$$\bar{\sigma}_c(t) = \frac{1}{N_f} \sqrt{\frac{1}{N_c(t)} \sum_{i=1}^{N_c(t)} [N_{c_i}(t) - N_f \bar{\mu}_c(t)]^2}, \tag{A.8}$$

quantifies the heterogeneity of cluster sizes, and forms an additional order parameter we will use to quantify the dynamics of filamentous clusters in the network.

## ACKNOWLEDGMENTS

We thank the members from the Center for Theoretical Biological Physics (CTBP) at Rice University and its memory-focused group. YE and MSC would like to thank Dr. Donald Kouri, Dr. Andre C. Barato, Dr. Kresimir Josic, and Jacob Tinnin from the University of Houston for their helpful suggestions and stimulating discussions. Moreover, we thank the Research Computing Data Core (RCDC) Center at the University of Houston for the computational resources. Ultimately, we thank the National Science Foundation for their funding support (CHE:1743392, PHY:2019745, OAC:1531814). Lastly, FJN thanks for his support by the Gatsby Charitable Foundation.

**Supporting Information for "Insights from Graph Theory on the Morphologies of Actomyosin Networks with Multilinkers"**

Yossi Eliaz[1,2], Francois Nedelec[3], Greg Morrison[1,2], Herbert Levine[2,4], and Margaret S. Cheung[1,2,5]*

[1]*Department of Physics, University of Houston, Houston, Texas 77204, USA*

[2]*Center for Theoretical Biological Physics, Rice University, Houston, Texas 77005, USA*

[3]*Sainsbury Laboratory, Cambridge University, Bateman Street, CB2 1LR Cambridge, UK*

[4]*Department of Physics, Northeastern University, Boston, Massachusetts 02115, USA*

[5]*Department of Bioengineering, Rice University, Houston, Texas 77030, USA*

*Corresponding author: mscheung@uh.edu

**Relevant Background Knowledge:**

### I. Dendritic spines in neurons

Neurons are brain cells that communicate using electrical and chemical signals through action potential signaling and neurotransmitter release. The two neurons are "in touch" but they do not physically touch one another. There is a junction, called a synapse, between two neurons where the presynaptic neuron signals the postsynaptic one. Dendritic spines are protrusions rising from postsynaptic neuron which form the receiver end of the synapse. Spine volume ranges between $10^{-3}$ μm³ and 1 μm³ [1], and each neuron may grow thousands of spines. Consistent with the Hebbian theory of learning and memory, spines expand during long-term potentiation (LTP) and shrink during long-term depression (LTD) [2]. These learning processes do not solely involve chemical and electrical signals, but also cellular mechanics. Experiments of high-resolution imaging have shown that LTP induction causes morphological changes in dendritic spines [3,4]. Besides the shrinking or expanding during LTD and LTP, the generation and destruction of spines have been observed [1]. The spine is rich with ABPs such as CaMKII crucial for maintaining the information and structural of a spine [5,6].

### II. CaMKII and the morphogenesis in the dendritic spine

The $Ca^{2+}$/calmodulin-dependent kinase II [7] (CaMKII) holoenzyme protein serves a dual role as a $Ca^{2+}$ signaling decoder and as a structural agent in directing calcium signals to change the makeup of actomyosin networks in a spine. It accounts for up to 2% of the mammalian brain proteome [7,8]. In vertebrates, there are four major CaMKII isoforms observed in over 40 different splice variants. Overall, these variants are genetically expressed in diverse tissue types. The two most prevalent isoforms in the brain are CaMKIIα and CaMKIIβ [9,10] and the number of the protein domains in these multimeric complexes varies from 12 to 14. Each unit is made up of several connected domains that serve distinctive functions [11]. Besides being a biochemical catalyzer and having autophosphorylation capabilities [12] in one domains, each monomeric unit is capable of binding actin filaments. To date, the highest actin-binding valency observed for CaMKII is six for the dodecamer (12-mer) CaMKIIβ isoform [10]. It makes CaMKII a rare kind of multivalent actin-binding protein [10,11,13]. It motivates the development of our hypothesis that ABP's valency affects the general morphology of actomyosin networks. It may elucidate a functional role for the multivalent association nature of the CaMKII, which regulates postsynaptic computational resources by changing the morphology of the dendritic spine [14,15]. CaMKII is essential for long-term potentiation [16] in a persistent strengthening of synapses. The morphological changes over time is called synaptic plasticity [11].

**Method sections:**

### I. Shape and mobility of fibers and solid objects in Cytosim

Filaments and multilinkers are described by vertices with three spatial coordinates each. A filament is an elastic, polar, and inextensible rod represented by a set of $p$ equidistant vertices $\{m_i\}_{i=1}^{p}$, where $m_1$ and $m_p$ are the minus and the plus ends of the filament, respectively. Filaments emulate filamentous actin (F-actin), whereas the skeleton of the multilinkers is treated as a solid object in Cytosim, i.e. a nondeformable set of vertices. The $k$ vertices $\{s_i\}_{i=1}^{k}$ of a solid object can be placed initially anywhere, but obey hard constraints on all pairwise distances ($|s_i - s_j| = d_{ij} = d_{ji}$); For each multilinker, we have distributed its vertices on the surface of a sphere. Hence the multilinker's binding entities that can possibly bind to filaments lay on the surface of a solid sphere, which effectively only has translational and rotational degrees of freedom.

### II. Interactions between objects in Cytosim

In Cytosim, any interaction between objects is expressed by a linearization $\mathbf{f}(\mathbf{r},t) = \mathbf{A}(t)\mathbf{r} + \mathbf{g}(t)$. The matrix $\mathbf{A}(t)$ and the vector $\mathbf{g}(t)$ contain the contributions from all the elementary interactions limited to the constant and linear terms of the Taylor expansion. For example, two points $\mathbf{r}_a$ and $\mathbf{r}_b$ from different objects ($a$ and $b$) can be connected by a link with Hookean stiffness $k$. The forces between the points are calculated as:

$$\mathbf{f}_a = -\mathbf{f}_b = k\left(1 - \frac{r_0}{|\mathbf{r}_b - \mathbf{r}_a|}\right)(\mathbf{r}_b - \mathbf{r}_a), \tag{1}$$

where $r_0 \geq 0$ is the resting length of the link. When a motor head is attached to a fiber, the motor's position is computed by a distance measured from a fixed reference point on the fiber ($\mathbf{x}_0$). The position is increased by $\delta = \tau v_{\max}^{\text{motor}} \left(1 - \frac{f}{f_{\text{stall}}}\right)$, where $v_{\max}^{\text{motor}} > 0$ is a constant real value representing the maximal speed of a motor head walking along a fiber. The value of $f$ is the load that the motor experiences projected along the direction of the filament on which the motor walks. The stall force $f_{\text{stall}}$ is the amount of force that is sufficient to stop the motor from moving. In general, Cytosim uses a force-dependent unbinding rate $k_{\text{off}} = k_0 \exp\left(\frac{|f|}{f_0}\right)$ to model dissociation from the fiber. Motors have constant binding $k_{\text{on}}^{\text{motor}}$ and unbinding rates $k_{\text{off}}^{\text{motor}}$. Namely, they work in the limit of constant unbinding rate ($f_0 \rightarrow \infty$). In addition, we set $v_{\max}^{\text{motor}} = 0.2$ μm s$^{-1}$ and $f_{stall} = 6$ pN. The bivalent crosslinker and multilinkers have the unbinding and binding rates: $k_{\text{on}}^{\text{linker}}$ and $k_{\text{off}}^{\text{linker}}$. The interaction of a linker and filaments is assumed to be Hookean with zero resting length and a stiffness of 50 pNμm$^{-1}$.

### III. Excluded volume interactions between multilinkers in Cytosim

Cytosim incorporates excluded volume and "steric" effects between objects. Generally, it supports both attractive and repulsive forces as the interaction is taken as piecewise linear force:

$$f(d) = \begin{cases} k_+(d - d_0), & d \leq d_0 \\ 0, & \text{else} \end{cases}, \qquad (2)$$

where $d$ is the distance between two interacting elements, $d_0 = 5$ nm is their equilibrium distance, and $k_+ = 500$ pNμm$^{-1}$ is the repulsion stiffness.

### IV. Parameters in the simulations

Our simulations have multiple parameters. Some of parameters, e.g., the valency of multilinkers, are varied and set up before the execution of the simulations and others are fixed across all the set of the simulations. Overall this study involved about 1,500 simulations of 600 seconds trajectories at different concentration of motors, linkers, multilinkers and filaments. All parameters are summarized in Table S.I.

TABLE S.I. System configuration parameter values for the computational model, with, when applicable, the set values that has been explored. In all simulations, these parameters are held constant from start to end, and they are only varied between simulations

| Parameter | Description | Values |
|---|---|---|
| $v$ | Valency of a multilinker | {2,3,4,5,6,7} |
| $k_{\text{multilinker}}$ | Hookean stiffness of a multilinker binding entity | 200 pNμm$^{-1}$ |
| $k_{\text{crosslinker}}$ | Hookean stiffness of a crosslinker | 250 pNμm$^{-1}$ |
| $k_{\text{off}}^{\text{linker}}$ | Unbinding rate of a multilinker or a crosslinker binding entity | 0.1 s$^{-1}$ [17] |
| $k_{\text{on}}^{\text{linker}}$ | Binding rate of a multilinker or a crosslinker binding entity | 5 s$^{-1}$ [18] |
| $k_{\text{off}}^{\text{motor}}$ | Unbinding rate of a motor binding entity | 0.1 s$^{-1}$ [19] |
| $k_{\text{on}}^{\text{motor}}$ | Binding rate of a motor binding entity | 10 s$^{-1}$ [20] |
| $N_{\text{motors}}$ | Number of motors | {10,250,500,1000} |
| $N_{\text{filaments}}$ | Number of filaments | {250,500,1000} |
| $N_{\text{crosslinker}}$ | Number of α-actinin crosslinkers | {10,250,500,1000} |
| $V_{box}$ | Volume of the simulation box | 1μm$^3$ |
| $N_{\text{multilinkers}}$ | Number of multilinkers | {250,500,1000} |

## V. Complementarity list of network-theory measures and order parameters

Node centrality has been useful in better understanding the structure of complex networks on the level of individual nodes [21,22]. There are many possible measures of centrality found in the literature, each measuring the 'importance' of a node in a slightly different way. This follows the convention that a graph $\mathbf{G} = (V, E)$ has a set of nodes V and edges E and the adjacency matrix of $\mathbf{G}$ is $\mathbf{A}$.

*Eigenvector centrality*: One example of node centrality order parameter is the eigenvector centrality of node $i$, which is defined by the $i$-th element of the eigenvector $\mathbf{v}$ defined by:

$$\mathbf{A}\mathbf{v} = \lambda_{\text{max}}\mathbf{v}, \qquad (3)$$

where $\lambda_{\text{max}} \geq \lambda_1 \geq \cdots \geq \lambda_{|V|}$ is the maximal eigenvalue of the symmetric adjacency matrix $\mathbf{A}$ of the graph $\mathbf{G}$. Eigenvector centrality tends to be higher for nodes that are connected to other nodes with high centrality. Then, the average eigenvector centrality is simply:

$$E_{\lambda_{\text{max}}}(t) = \frac{1}{|V(t)|}\sum_{1 \leq i \leq |V(t)|} \lambda_i. \qquad (4)$$

*Average neighbor degree*: The average neighbor degree of a node is another common measure for the centrality of a node $v$ and is derived from $v$'s neighborhood. The nearest neighbors of a node $v$ is defined as $\mathcal{N}(v) = \{u \mid (u,v) \in E\}$ and the neghibor degree of $v$ is given by:

$$k_{nn}(v) = \frac{1}{|\mathcal{N}(v)|}\sum_{u \in \mathcal{N}(v)} k_u, \quad (5)$$

where $k_u$ is the degree of the node $u$. Like eigenvector centrality, neighbor degree centrality is higher for nodes that are connected to other central nodes, but unlike eigenvector centrality longer paths are not explicitly accounted for. Then we define an order parameter as the mean average neighbor degree:

$$k_{nn}(t) = \frac{1}{|V|}\sum_{v \in V} k_{nn}(v). \quad (6)$$

*Betweenness centrality*: The betweenness centrality of a node (referred to as "betweenness") measures how central a node is by how often it acts as a `bridge' between other pairs of nodes (sketched in **Fig. 2** in the main text). This is quantified by counting the number of shortest paths in which the node participates:

$$\beta(v \in V) = \sum_{s,s' \in V^*} \frac{g(s,s'|v)}{g(s,s')}, \quad (7)$$

where $V^* = V \setminus \{v\}$ is the set of nodes excluding node $v$, the function $g(s,s')$ counts the number of geodesics (shortest paths) between node $s$ and node $s'$, and $g(s,s'|v)$ is the number of shortest path between node $s$ and $s'$ such that $v$ lies along those shortest paths.

## VI. Potential of Mean Force of Actomyosin Networks

The potential of mean force [23] is the natural logarithm of a probability density function (PDF) defined on a degree of freedom in the system. We let $P(\xi)$ be such a probability density function (PDF) of a parameter $\xi \in [0,1]$. Since $\xi$ is an order parameter, by definition, there exists a mapping from the configuration space of the system onto the segment [0,1]. Subsequently, we can write $\xi$ to be this function $\xi(\mathbf{x}(t))$ which can be evaluated at each time point based on the configuration of the system. Along a trajectory, the probability density function of $\xi$ is given by:

$$P(\xi) = \langle \delta(\xi - \xi(\mathbf{x})) \rangle_{\text{trajectory}} \stackrel{\text{def}}{=} \frac{Q(\xi)}{\Sigma_\xi Q(\xi)}, \qquad (8)$$

where $Q(\xi)$ is the partition function to find the system in a state at which the order parameter of interest has the value $\xi$. The entropy $S(\{\xi\})$ [24] [25] on the probability density function of $\xi$ is given by:

$$S(\{\xi\}) = -K_B \sum_{\xi'} P(\xi') \ln P(\xi') = k_B \sum_{\xi'} P(\xi') \, \text{I}(\xi'), \qquad (9)$$

where $\text{I}(\xi) \stackrel{\text{def}}{=} -k_B \ln P(\xi)$ is the Shannon information (self-information) as a function of $\xi$.

If we let $k_B$ be unity, both the entropy and the information measure have the nat (natural unit of information) units [26]. The Shannon information function is defined:

$$\text{I}(\xi) \stackrel{\text{def}}{=} -k_B \ln P(\xi) \stackrel{\text{def}}{=} \ln \left( \frac{\Sigma_{\xi'} Q(\xi')}{Q(\xi)} \right). \qquad (10)$$

We used the python KernelDensity estimator the Scikit Python library [27] to evaluate the self-information $\text{I}(\xi)$ via the negative logarithm of the PDF $P(\xi)$. The self-information is also related to the potential of mean force (PMF), noted $\mathcal{U}(T, \xi)$ [23,28] for which the following holds by definition:

$$\mathcal{U}(T, \xi) \stackrel{\text{def}}{=} T\text{I}(\xi) \stackrel{\text{def}}{=} -k_B T \ln P(\xi) \stackrel{\text{def}}{=} k_B T \ln \left( \frac{\Sigma_{\xi'} Q(\xi')}{Q(\xi)} \right). \qquad (11)$$

**Results:**

### I. Polymer Physics Order Parameters

The computed order parameters from polymer physics on the actin vertices yield not a very meaningful behavior in the time signal and the network representation in Cytosim is too sparse compared with our previous finer-grained representation of actomyosin [29]. However, it is possible to draw meaningful interpretation of the effect of motor on the network. The values of all three order parameters from the moment of inertia have similar pattern to the cumulative distribution function (CDF) in **Fig. S1** which is high concentrated around a single value along all the trajectories and the variation with higher motor is higher.

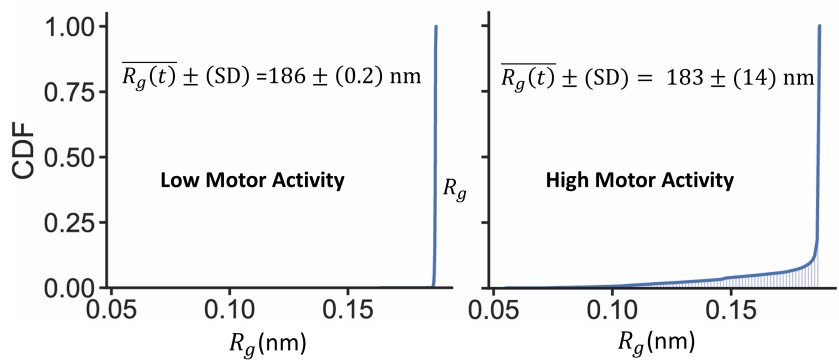

**FIG. S1**. Cumulative distribution functions of the radius of gyration $R_g$ at low and high motor activity.

The fluctuations in these order parameters reveal the stress relaxation the system undergoes through, and its response at different ABP-filament concentration ratios. Using spectral and fluctuation analysis, we evaluate the differences [30] at high versus low motor activity. The fluctuation in the distribution of filament vertices in space is a proxy for the temporal dynamics of the actomyosin networks since the filaments are the scaffold of the network. Using the power spectral density functions (PSD) [31] of the three polymer physics order parameters $R_g$, $\Delta$, and S defined in Eq. A.(3-5) in the appendix: are compared for high and low motor-to-filament ratios in **Fig. S2**. Each of these three order parameters measures a distinctive mesoscopic reshaping mode (e.g., whether the filaments slides, rotate with respect to the center of mass of the network) of the actomyosin network. PSD of the shape parameter, S, is a mesoscopic assessment of filament sliding modes (Fig 2 (a-b)). The PSD of the asphericity ($\Delta$) is a proxy for the rotational modes of filaments (Fig 2(c-d)). The PSD of the radius of gyration ($R_g$) order parameter corresponds to mesoscopic expansion-contraction modes (**Fig. S2**(e-f)), The filament sliding mode in **Fig. S2**(a-b) is more dominant than the rotational mode in (**Fig. S2**(c-d) because The PSD of the former is one order of magnitude (10dB) higher than the PSD of the latter. When comparing the PSD at the two low and high conditions of motor-filament-crosslinker-multilinker concentrations, (1:100:100:100) vs. (1:1:1:1), all the three PSD in the case of high motor-filament ratio is two order of magnitudes (20dB) greater than those of low motor-filament ratio.

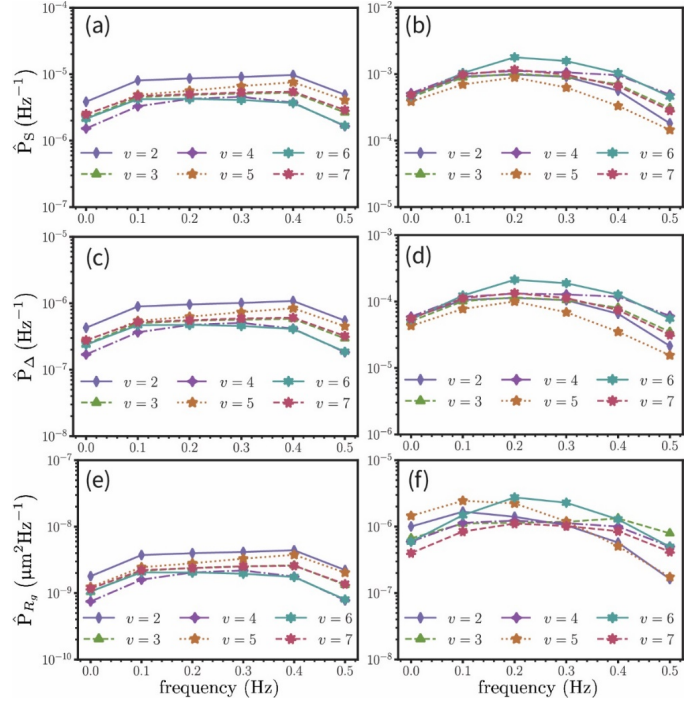

**FIG. S2**. Power spectral analysis of mesoscopic polymer order parameters. Low and high motor-filament-crosslinker-multilinker ratios, (1:100:100:100) and (1:1:1:1) are examined in systems with multilinkers of various valencies $v \in \{2,3,4,5,6,7\}$. (a), (c) and (e) show the PSD from the systems with low motor-filament ratios, while (b), (d), and (f) show the PSD from the systems with high motor-filament ratios. We compared the PSD for the shape parameter (S), asphericity ($\Delta$), and the radius of gyration, $R_g$.

## II. Gelation of the actomyosin network by actin-binding proteins

The order parameters from gelation theory compute the clusters based on the connectivity of filaments through ABPs, and thus resemble the parameters from percolation theory. The largest cluster of filaments is a key parameter in percolation theory applied to finite networks as in the case of our actomyosin networks. **Fig. S3**(a-b) shows the normalized largest cluster size over time, $\bar{N}_{\max}(t)$, from the Appendix. This demonstrates how multilinkers with higher valency promote larger clusters regardless of motor concentration. Likewise, **Fig. S3**(c-d) show how the normalized standard deviation of the cluster size, $\bar{\sigma}_c(t)$ defined in the Appendix, is larger for systems with multilinkers of higher valency because the higher the valency of the multilinkers is in the system size, the greater the size cluster of the largest cluster size, which skews the standard deviation. This is the reason why $\bar{\sigma}_c(t)$ increases in the presence of multilinkers with higher valencies at both high and low motor to filament concentration ratios.

The total number of clusters formed by multilinkers with varying valency depends on the content of motor concentration. At a low motor content in **Fig. S3**(e), the systems with lower valency shows greater number of total clusters. In contract, at a high motor content, all systems show a similar total number of clusters regardless of the valency in multilinkers (**Fig. S3**(f)).

**Fig. S3**(g) shows the cumulative distribution functions (CDF) of the average cluster size, $\bar{\mu}_c$ at a high and a low motor content, both in the presence of multilinkers with varying valencies. Multilinkers with valency $v = 2$ and $v = 3$ create smaller clusters on average at low motor concentration than at a high motor concentration. For systems with multilinkers of $v = 4$, both systems with high and low motor to filament ratios have a similar distribution of $\bar{\mu}_c$. Finally, at $v = 5, v = 6$, and $v = 7$, it appears that high motor activity promotes small cluster. It is because the largest cluster contains more than 60% of the filaments in the system, and this in turn skews the distribution of the average cluster size.

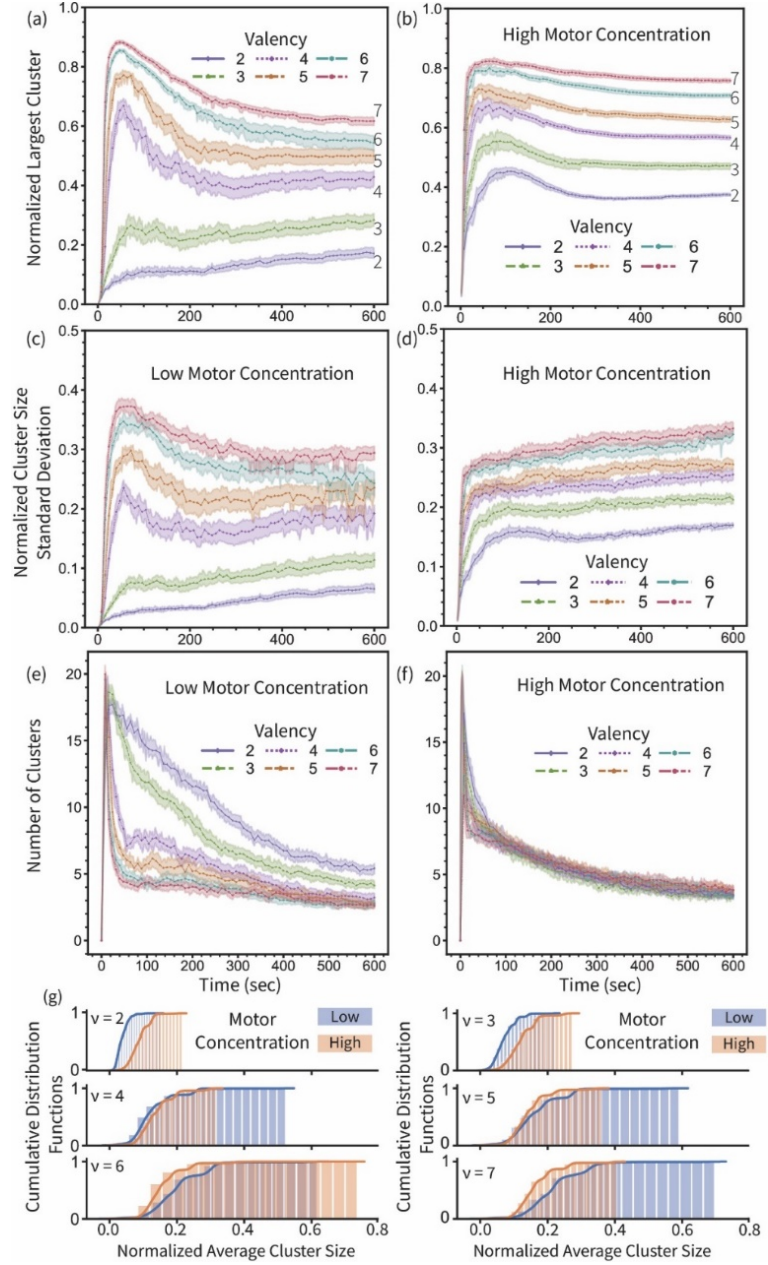

**FIG. S3**. Temporal evolution of gelation formed by filaments and crosslinkers, multilinkers, and motors. The condition of "low motor concentration" corresponds to motor-filament-multilinker-crosslinker ratios of 1:100:100:100 in a box of 1000 filaments as in (a, c, e). The condition of "high motor concentration" corresponds to motor-filament-multilinker-crosslinker ratios of 1:1:1:1 in a box of 1000 filaments as in (b, d, f). (a-b) The normalized largest cluster is plotted against time. (c, d) The standard deviation of the normalized cluster size $\bar{\sigma}_c(t)$. (e, f) The number of clusters against time. (g) The cumulative distribution functions (CDF) of the normalized average cluster size order parameter $\mu_c(t)$. The profiles from each panel show six valency conditions in multilinkers and each profile was averaged from 30 independent trajectories with a confidence interval of CI = 90%.

**Input configuration file of a multilinker Cytosim simulation:**

```
set simul system
{
 time_step              = 0.004;
 viscosity              = 0.5;
 random_seed            = 3158880834;
 steric                 = 1, 500, 100;
}

set space cell
{
 property_number        = 1;
 shape                  = square;
 dimensions             = 1 1 1;
}

set fiber filament
{
 property_number        = 1;
 rigidity               = 0.075;
 segmentation           = 0.1;
 viscosity              = 0.5;
 confine                = 1, 100, first;
 display                = (line=0.5, 1; color=orange;);
}

set hand motor
{
 property_number        = 1;
 binding                = 10, 0.05;
 unbinding              = 0.1, inf;
 display                = (size=2; color=green;);
 activity               = move;
 stall_force            = 6;
```

```
 unloaded_speed        = 0.2;
}

set hand binder
{
 property_number       = 2;
 binding               = 5, 0.0175;
 unbinding             = 0.1, inf;
 display               = (size=2; color=blue;);
}

set couple crosslinker
{
 property_number       = 1;
 hand1                 = binder;
 hand2                 = binder;
 stiffness             = 250;
 diffusion             = 10;
 fast_diffusion        = 1;
}

set couple complex
{
 property_number       = 2;
 hand1                 = motor;
 hand2                 = motor;
 stiffness             = 250;
 diffusion             = 10;
 fast_diffusion        = 1;
}

set hand binder_multivalence
{
 property_number       = 3;
 binding               = 5, 0.006;
 unbinding             = 0.1, inf;
```

```
    display             = (size=2; color=blue;);
}

set single grafted
{
 property_number     = 1;
 hand                = binder_multivalence;
 stiffness           = 200;
}

set solid blob
{
 property_number     = 1;
 viscosity           = 0.5;
 steric              = 10, 0;
 confine             = 1, 100, cell;
 display             = (style=7; coloring=0; color=0x88888888;);
}
```